\documentclass[%
 reprint,
 amsmath,amssymb,
 aps,
]{revtex4-2}

\usepackage{graphicx}
\usepackage{dcolumn}
\usepackage{bm}
\usepackage{appendix}
\usepackage{xcolor}
\begin{document}

\preprint{APS/123-QED}

\title{Foundation of statistical mechanics under even more experimentally realistic conditions}
\author{M. R. Passos}
\author{Thiago R. de Oliveira}
\email{troliveira@id.uff.br}
\affiliation{Instituto de Física, Universidade Federal Fluminense, Av. Litor{\^a}nea s/n, Gragoatá 24210-346, Niterói, RJ, Brazil
}

\date{\today}

\begin{abstract}

Understanding how macroscopic systems exhibit irreversible thermal behavior has been a longstanding challenge, first brought to prominence by Boltzmann. Recent advances have established rigorous conditions for isolated quantum systems to equilibrate to a maximum entropy state, contingent upon weak assumptions. These theorems, while powerful, apply for a sudden quench. However, natural processes involve finite-time perturbations or quenches, which raises a crucial question: Can these systems still equilibrate under more realistic, finite-time dynamics? In this work, we extend the established results to account for finite-time quenches, demonstrating that even under finite-time perturbations, the system will equilibrate provided it populates many significant energy levels.  While the mathematical proof is more intricate than in the instantaneous case, the physical conclusion remains the same: sufficient perturbation leads to equilibration. Our results provide a broader and more physically realistic framework for understanding thermalization in isolated quantum systems.

\end{abstract}

\maketitle
\section{Introduction}
Thermodynamics, one of the most successful theories in physics, offers profound insights into macroscopic phenomena without requiring detailed knowledge of microscopic structures or laws. However, it leaves a crucial open question: how do the microscopic laws of quantum mechanics give rise to the macroscopic laws of thermodynamics, particularly the tendency toward thermal equilibrium? Statistical physics aims to bridge this gap, though its foundations have been debated since its inception. A recent perspective focuses on isolated quantum systems that evolve unitarily, showing that while the full system state never reaches equilibrium, many observables do because, for large systems, fluctuations become negligible. This phenomenon termed probabilistic convergence or equilibration on average, has been the subject of rigorous theoretical investigation and some experiments \cite{Gogolin16}.

The cornerstone result in this field is a theorem proving that, under weak assumptions, most isolated quantum systems thermalize after a perturbation \footnote{The recent results were obtained in \cite{Reimann08, Linden08}. Later, it was found out that von Neumann already obtained similar results \cite{Goldstein10}. A not-so-recent but also important and usually forgotten reference is \cite{Tasaki98} }. While some assumptions of the original theorem were taken into account in later generalizations, as degeneracies \cite{Reimann12, Short12}, a critical one remains: that the system perturbation is instantaneous; a sudden quenche. In practice, perturbations take a finite amount of time, and their impact on the system’s equilibration must be carefully studied. Clearly, one expects that finite but very fast quenches will still lead to equilibration while very slow ones will not; actually, they will not even be able to take the system out of equilibrium. Here, we rigorously show that, in fact, this is the case, extending the equilibration theorems to systems that undergo finite-time quenches. Interestingly, the results are similar, although the proof is more intricate. Thus we offer
a more comprehensive understanding of equilibration under realistic experimental conditions.

Let us first briefly review the results for a sudden quench. One considers a system initialized in an eigenstate $|E_m^i\rangle$ of the Hamiltonian $H_{i}$, which is then suddenly changed or quenched to $H_{\infty}=\sum_n E_n |E_n \rangle \langle E_n| $. The time evolution of the system is described by the quantum state $|\Psi(t)\rangle$, namely,   
\begin{equation}
|\Psi(t)\rangle = \sum_{n} d_{n}(0^{+}) e^{-i E_{n}t}|E_n\rangle,     
\end{equation}
where $d_{n}(0^{+}) \equiv \langle E_n | E_m^i\rangle$. This eigenbasis is chosen such that in case there is degeneracy $|E_m^i\rangle$ has non-zero overlap with only one eigenstate $|E_n\rangle$  for each energy eigenvalue $E_{n}$. 

After the quench, the expectation value of the observable $A$ will oscillate around an average value, the infinite-time average $\overline A$. Although the fluctuation $\Delta A(t) = \langle A \rangle(t) - \overline{A} $ never goes to zero, it can become very small most of the time as the system size increases. In this case, the system is equilibrated for all practical purposes. One way to quantify this is by the infinite-time average fluctuation. The cornerstone result is the upper bound \cite{Short12}
\begin{equation}
\label{fundamentalquench}
\overline{\Delta A(t)^2} \leq \frac{||A||^2}{d_{eff}},    
\end{equation}
with the infinite-time average of a time-dependent quantity $f(t)$ defined by $ \overline{f(t)} \equiv \lim_{T \rightarrow \infty} \frac{1}{T} \int_{0}^{T}d\tau f(\tau)$ and  $\langle A\rangle(t) \equiv tr(\rho(t)A)$, $\bar{A} \equiv tr(\bar{\rho}A)$. Here $\bar{\rho}$ is the average state $\bar{\rho} \equiv \overline{|\Psi(t)\rangle\langle\Psi(t)|}$. The denominator on the rhs of \eqref{fundamentalquench} is the effective dimension $d_{eff} \equiv 1 / tr(\bar{\rho}^2) = \sum_{n}|d_{n}(0^{+})|^4$ and the symbol $\| \bullet \|$ means the operator norm. The only assumption about the systems is that the energy gaps of $H_{\infty}$ are non-degenerate, but they can be relaxed \cite{Short12}. This result is very strong since, under weak assumptions, it gives a sufficient condition for an initial out-of-equilibrium state to equilibrate; it needs a large $d_{eff}$ \footnote{In fact, $d_{eff}$ has to increase faster than $||A||$, which is usually the case for local extensive observables as the magnetization. But there are always non-local observables that will not equilibrate}. And $d_{eff}$ quantifies how much the initial state spread over the spectrum $H_{\infty}$. Moreover, for local $H$, the distance between the energy levels is exponentially small with the system size $N$. Therefore it is almost impossible to prepare an initial state with small $d_{eff}$. Note that the average state $\bar{\rho}$ is the unique state that maximizes the entropy, given the conserved quantities (populations) \cite{Gogolin16} and, therefore, a kind of thermal state. In sum, Eq. 2 would explain thermalization under experimental realistic conditions as stated in the original work \cite{Reimann08}.

One relevant open question is what happens if the perturbation is not instantaneous. We consider that the initial Hamiltonian $H_i$ is continuously
changed to $H_i + \Delta H$: $H(t) = H_i + \lambda(t) \Delta H$ with $\lambda(t)$ a function that varies from $0$ to $1$. The quench case has $\lambda(t)$ as a step function at $t=0$. Thus, we generalize \eqref{fundamentalquench} to a broader class of systems, with a Hamiltonian that depends continuously on time. Actually, we assume that the Hamiltonian $H(t)$ may change arbitrarily from an initial time to a finite time $t^{*}$, then $H(t)$ approaches a constant operator value $H_{\infty}$, namely,
\begin{equation}
\label{limitconstant}
\lim_{t\rightarrow \infty} H(t) \equiv H_{\infty} = H_i + \Delta H.    
\end{equation}
The only condition on the time evolution is imposed for $t \geq t^{*}$:
\begin{equation}
\label{decaying}
\| H(t) - H_{\infty} \| \leq \frac{K}{t^{2 + \epsilon}}.
\end{equation}
This is necessary to guarantee the convergence of $\overline{\Delta A(t)^{2}} $. Here $K$ is a positive constant and $\epsilon > 0$. From now on, we denote $\delta H(t) \equiv H(t) - H_{\infty}$.

To illustrate our assumptions \eqref{limitconstant} and \eqref{decaying}, we first consider 
a linear interpolation, such that $\lambda(t)$ is defined by  
\begin{equation}
\label{lambda}
\lambda(t) \equiv 
\begin{cases}
    t/t^{*}, \text{  } 0 \leq t \leq t^{*}, \\ 
    1, \text{  } t \geq t^{*}.
\end{cases}
\end{equation}
In this case, $\lim_{t \rightarrow \infty} H(t) = H_{0} + \Delta H = H_{\infty}$ and $\| \delta H(t) \| \equiv 0  \leq K/ t ^{2+\epsilon}$ (for $t \geq t^{*}$), thus fulfilling our assumptions. Actually, the quench is a special case of \eqref{lambda} when $t^{*} \rightarrow 0^{+}$.
As another example, consider now $\lambda(t)$ defined by
\begin{equation}
\label{lambda2}
\lambda(t) \equiv 
\begin{cases}
    t/t^{*}, \text{  } 0 \leq t \leq t^{*}, \\ 
    (t^{*}/t)^3, \text{  } t \geq t^{*}.
\end{cases}
\end{equation}
Now $\lim_{t \rightarrow \infty } H(t) = H_{0}$ and $\| \delta H(t) \| = \| \Delta H \| (t^{*}/t)^3 \leq K/ t ^{2+\epsilon}$ (for $t \geq t^{*}$) which is in agreement with \eqref{limitconstant} and \eqref{decaying}, this latter with $K = (t^{*})^3 \| \Delta H \|$ and $\epsilon = 1$. We have also assumed that $ \Delta H$ is bounded. In fact, $\lambda(t)$ defined in \eqref{lambda2} is different from that defined in \eqref{lambda}, since in \eqref{lambda2} the external perturbation is turned on and off. Actually, we may think of it as an external perturbation that is turned on and then slowly fades away. 

The generalization of \eqref{fundamentalquench} involves the description of the quantum state of the system in the interaction picture, where the operators evolve by the \textit{free} time-independent part of $H(t)$, while the states evolve by the \textit{non-free} time-dependent part of $H(t)$. Here, in a way that may seem counterintuitive, we write $H(t) = H_{\infty} + \delta H(t)$ and the time evolution corresponding to $H_{\infty}$ is imparted to operators while that corresponding to the difference $\delta H(t)$ is imparted to state vectors. Thus, we denote the quantum state of the system in the interaction picture by $|\Psi_{I}(t)\rangle \equiv e^{i H_{\infty}t}|\Psi(t)\rangle$, in which no subscripts are used for the Schr\"odinger picture. Operators in the interaction picture also carry the subscript like $A_{I}(t) \equiv e^{i H_{\infty}t}Ae^{-i H_{\infty}t}$.
The reason for such a choice is that it makes that the Hamiltonian $\delta H_{I}(t) \equiv e^{i H_{\infty}t}\delta H (t)e^{-i H_{\infty}t}$ vanishes and, as a consequence, the quantum state in the interaction picture $|\Psi_{I}(t)\rangle$ approaches a well defined infinite-time limit quantum state  $|\Psi_{I}(\infty)\rangle \equiv lim_{t \rightarrow \infty} |\Psi_{I}(t)\rangle$, which is demonstrated below. This last point is not trivial; indeed, we observe that in the Schrodinger picture, the same does not happen, but instead, $|\Psi(t)\rangle$ in general does not approach any vector in the space, i.e., $lim_{t\rightarrow \infty} |\Psi(t)\rangle$ does not exist.   
A consequence of this preceding fact is the existence of an interaction-picture average state $\overline{\rho_{I}}$, which allows us to generalize the fluctuation theorem \eqref{fundamentalquench}. Indeed, it is important to mention that while the quantum state in the Schrodinger picture never approaches a well-defined state, the quantum state in the interaction picture always approaches a well-defined state. Even in the quench of the Hamiltonian the quantum state in the interaction picture approaches $\lim_{t \rightarrow \infty}|\Psi_{I}(t)\rangle = \lim_{t \rightarrow \infty}e^{iH_{\infty}t}|\Psi(t)\rangle = |\Psi(0)\rangle$. In this last case, we may think that $|\Psi_{I}(t)\rangle$ reaches its limit value $|\Psi(0)\rangle$ instantaneously at $t = 0$.

We also consider the same quench hypotheses that $H_{\infty}$ may have a degenerate energy spectrum, but the energy gaps of its spectrum are non-degenerate. We write it as
\begin{equation}
\label{hindecom}
H_{\infty} = \sum _{n,k} E_{n}|E_{n,k}\rangle \langle E_{n,k}|,      
\end{equation}
where $E_{n}$ are distinct energy eigenvalues and $\{ |E_{n,k}\rangle \}$ form an orthornormal eigenbasis of $H_{\infty}$, that is, $\langle E_{n,k}|E_{m,l} \rangle = \delta_{nm}\delta_{kl}$. The index $k$ accounts for possible energy degeneracies. Furthermore, the coefficients of $|\Psi_{I}(t)\rangle$ with respect to the eigenbasis of  $H_{\infty}$ are written as $c_{n}^{k}(t) \equiv \langle E_{n,k}| \Psi_{I}(t)\rangle$ while in the Schr\"odinger picture we write $d_{n}^{k}(t) \equiv \langle E_{n,k}| \Psi(t)\rangle$.
Let us also assume that the system is initially prepared in an eigenstate of $H(0)$, that is, $|\Psi(0)\rangle = |E_{m}^{i}\rangle$, where $H(0)|E_{m}^{i}\rangle = E_{m}|E_{m}^{i}\rangle$.

\section{Theorem about Quantum Fluctuations}
 Let $\rho(t)\equiv |\Psi(t)\rangle \langle \Psi(t)|$ be the quantum state of the system. We define the interaction-picture average state of the system $\bar{\rho}_{I}$ by the expression
\begin{equation}
\label{averagestate}
 \bar{\rho}_{I} \equiv \overline{e^{iH_{\infty}t}\rho(t)e^{-iH_{\infty}t}}.
\end{equation}
 We state the following result: for a quantum observable $A$,
\begin{equation}
\label{fundamentalresult}
 \overline{\Delta A(t)^{2}} \leq \frac{||A||^2}{d_{eff}},     
\end{equation}
where $\langle A\rangle(t) = tr(\rho_{I}(t)A_{I}(t))$ and we define $\bar{A} \equiv tr(\bar{\rho}_{I}A)$. Here $d_{eff} = 1/tr(\bar{\rho}_{I}^2)$ is the effective dimension. We observe that the preceding definition of $\bar{A}$ indeed equals to the infinite-time average of $\langle A\rangle(t)$, because $\overline{\langle A\rangle(t) - \bar{A}} = 0$, with $\bar{A} \equiv tr(\bar{\rho}_{I}A) $ (see Appendix \ref{AppD}). Another important remark is that $d_{eff} = 1/tr(\bar{\rho}_{I}^2)$ is different from that in \eqref{fundamentalquench}. In fact, the effective dimension in \eqref{fundamentalquench} is related to the projection of the initial state of the system on the $H_{\infty}$ basis, that is, if the initial state significantly spreads over the $H_{\infty}$ basis and $d_{eff}$ is large, then the system reaches an equilibrium. On the other hand, in \eqref{fundamentalresult} $d_{eff}$ has to do with the projection of $|\Psi_{I}(\infty)\rangle$ over the $H_{\infty}$ basis, that is, it is not simple to determine what is the role of the initial state in the equilibration, because it is subjected to the dynamics generated by $\delta H_{I}(t)$. In other words, it is not a matter of only determining its projection on the $H_{\infty}$ basis because the dynamics also contribute to spreading the initial state over the $H_{\infty}$ basis in a way that cannot be exactly predicted.
 
To prove the existence of \eqref{averagestate} and the validity of \eqref{fundamentalresult}, we begin by considering that in the interaction picture the limit $c_{n}^{k}(\infty) \equiv \lim_{t \rightarrow \infty} c_{n}^{k}(t) =  \lim_{t \rightarrow \infty} \langle E_{n,k}| \Psi_{I}(t)\rangle$ exists. 
Indeed, for $\tau \geq t \geq t^{*}$ we write   
\begin{equation}
\label{dev1}
|c_{n}^{k}(\tau) - c_{n}^{k}(t)| = |\langle E_{n,k}|\Big{(}\mathcal{U}_{I}(\tau,t) -\mathbf{1}\Big{)}\mathcal{U}_{I}(t,0)|E_{m}^{i}\rangle|,
\end{equation}
where $\mathcal{U}_{I}(t_{2},t_{1})$ is the time evolution operator in the interaction picture, namely, $|\Psi_{I}(t)\rangle = \mathcal{U}_{I}(t,0)|\Psi_{I}(0)\rangle$ (see \eqref{evolutionui}), which possesses the group property $\mathcal{U}_{I}(t_{3},t_{1}) = \mathcal{U}_{I}(t_{3},t_{2})\mathcal{U}_{I}(t_{2},t_{1})$. 

We expand the operator $\mathcal{U}_{I}(\tau,t) -\mathbf{1}$ in a Dyson series as
\begin{multline}
\label{dev2}
\mathcal{U}_{I}(\tau,t) -\mathbf{1} = \mathcal{T} \Bigg{(} \sum_{n=1}^{\infty} \frac{(-i)^n}{n!} \int^{\tau}_{t} dt_{1} \delta H_{I}(t_{1}) ...\\ \int_{t}^{\tau}dt_{n} \delta H_{I}(t_{n}) \Bigg{)},    
\end{multline}
where $\mathcal{T}$ is the time-ordering operator and $\delta H_{I}(t) \equiv e^{iH_{\infty}t} \delta H(t) e^{-iH_{\infty}t}$. An upper-bound on $||\mathcal{U}_{I}(\tau,t) -\mathbf{1}||$ can be obtained from \eqref{dev2} by observing that the hypothesis \eqref{decaying} and triangle inequality imply that
\begin{equation}
\label{dev3}
\Bigg{|}\Bigg{|} \int_{t}^{\tau} dt' \delta H_{I}(t') \Bigg{|}\Bigg{|} \leq \frac{K}{(1+\epsilon)}\Bigg{(}\frac{1}{t^{1+\epsilon}} - \frac{1}{\tau^{1+\epsilon} } \Bigg{)} 
\end{equation}
and we have considered that $||\delta H_{I}(t)|| = ||\delta H(t)||$.
The norm of the series on the rhs of \eqref{dev2} then furnishes
\begin{equation}
\label{dev4}
 ||\mathcal{U}_{I}(\tau,t) -\mathbf{1}|| \leq e^{\frac{K}{(1+\epsilon)}\Big{(}\frac{1}{t^{1+\epsilon}} - \frac{1}{\tau^{1+\epsilon} } \Big{)}} - 1,  
\end{equation}
which follows from triangle inequality, submultiplicativity\footnote{$||A B|| \leq ||A|| ||B||$} and \eqref{dev3} (see Appendix \ref{appA}). As we can see from \eqref{dev4} by taking the limit $\tau \rightarrow \infty$ on both sides of this equation, the time evolution operator in the interaction picture $\mathcal{U}_{I}(\infty,t)$ approaches the identity. This is made possible because the generator of the time evolution in the interaction picture $\delta H_{I}(t)$ vanishes accordingly with \eqref{decaying} for $t \geq t^{*}$.

The result obtained in \eqref{dev4} when applied to the rhs of \eqref{dev1} imediatelly provides
\begin{equation}
\label{dev5}
|c_{n}^{k}(\tau) - c_{n}^{k}(t)| \leq e^{\frac{K}{(1+\epsilon)}\Big{(}\frac{1}{t^{1+\epsilon}} - \frac{1}{\tau^{1+\epsilon} } \Big{)}} - 1,   
\end{equation}
which also follows from submultiplicativity and $||\mathcal{U}_{I}(t,0)||=1$. Finally we take the limit $\tau \rightarrow \infty$ on both sides of \eqref{dev5}, which leads to
\begin{equation}
\label{dev6}
|c_{n}^{k}(\infty) - c_n^{k}(t)| \leq e^{ \frac{K}{(1+\epsilon)t^{1+\epsilon} }}  - 1.
\end{equation}
Equation \eqref{dev6}  confirms the existence of the limit $c_{n}^{k}(\infty)$, as the difference $|c_{n}^{k}(\infty) - c_{n}^{k}(t)|$ diminishes as $t$ increases. This implies that the quantum state approaches a steady state over time in the interaction picture. Furthermore, we highlight that $\lim_{t \rightarrow \infty} |\Psi_{I}(t)\rangle \equiv |\Psi_{I}(\infty)\rangle = \sum_{n,k}c_{n}^{k}(\infty)|E_{n,k}\rangle$, but in general $\lim_{t \rightarrow \infty} |\Psi(t)\rangle = \lim_{t \rightarrow \infty} e^{-i H_{\infty}t}|\Psi_{I}(t)\rangle$ does not exist. Actually, the quantum state in the Schrodinger picture changes without approaching any state vector in space. Indeed, we observe that $ \lim_{t \rightarrow \infty} d_{n}^{k}(t) = \lim_{\rightarrow \infty} e^{i E_{n}t}c_{n}^{k}(t)$ does not exist, because the exponential oscillates indefinitely.

The average state in \eqref{averagestate} can be obtained by decomposing $\rho_{I}(t) = |\Psi_{I}(t)\rangle \langle\Psi_{I}(t)|$ in terms of an eigenbasis of $H_{\infty}$ and choosing this basis such that $|\Psi_{I}(\infty)\rangle$ has non-zero overlap with only one eigenstate $|E_{n,k}\rangle$ (denoted simply by $|E_{n}\rangle$) for each energy eigenvalue, which provides (see Appendix \ref{appB})
\begin{equation}
\label{dev7}
\overline{\rho}_{I} = \sum_{n} |c_{n}(\infty)|^2 |E_n\rangle\langle E_n|.
\end{equation}

In order to demonstrate \eqref{fundamentalresult} we begin by writing $\langle A \rangle(t)$ and $\bar{A}$, namely,
\begin{equation}
\label{dev8}
\langle A\rangle(t) = \sum_{n,k}\sum_{m,l} c_{n}^{k}(t)^{*}c_{m}^{l}(t)e^{i(E_{n}-E_{m})t}\langle E_{n,k}|A|E_{m,l}\rangle
\end{equation}
and
\begin{equation}
\label{dev9}
\bar{A} = \sum_{n} |c_{n}(\infty)|^2 \langle E_n|A|E_n\rangle,
\end{equation}
where the choice of basis aforementioned is to be considered. Expressions \eqref{dev8} and \eqref{dev9} then provide
\begin{multline}
\label{dev10}
 \overline{\Delta A(t)^2} =  \sum_{n \neq m} \sum_{p \neq q} \overline{e^{i[(E_{n}-E_{m})-(E_{p}-E_{q})]t}} \times \\ 
 \times \big{(}c_{p}(\infty)^{*}A_{pq}c_{q}(\infty)\big{)}^{*}  \big{(}c_{n}(\infty)^{*}A_{nm}c_{m}(\infty)\big{)},
\end{multline}
where $A_{nm} = \langle E_n | A | E_m \rangle$.
Because we assume that there are no energy gap degeneracies in the spectrum of $H_{\infty}$, the latter expression reduces to
\begin{equation}
\label{dev11}
\overline{\Delta A(t)^2} = \sum_{n \neq m} |c_{n}(\infty)^{*}A_{nm}c_{m}(\infty)|^{2}
\end{equation}
and this latter leads to \eqref{fundamentalresult} (see Appendix \ref{AppC}). 

An important point here is that the choice of the power $t^{2+\epsilon}$ that we made in \eqref{decaying} (with $\epsilon > 0$) guarantees the convergence of the average value on the lhs of \eqref{dev10}. 
Another important point is that our result \eqref{fundamentalresult} is a particular case of \eqref{fundamentalquench} (see Appendix \ref{AppE}).

\section{Example}
 To illustrate our results, we consider a non-integrable spin chain of $N$ spins given by the Hamiltonian
\begin{multline}
H(t) =  J_{1}(t)\sum_{i = 1}^{N-1}(S_i^x S_{i+1}^x + S_i^y S_{i+1}^y + d S_i^z S_{i+1}^z) + \\
+ J_2(t)\sum_{i = 1}^{N-2} (S_i^x S_{i+2}^x + S_i^y S_{i+2}^y + d S_i^z S_{i+2}^z) +\\ + h_x S_i^x + h_z S_i^z + e S_1^x.
\end{multline}
Here $S^{x,y,z}$ are the Pauli spin matrices and $\hbar = 1$.
We use $h_z=0$, $h_x=0.2$, $d=0.5$, $e=0.2$. The perturbation is to linearly turn on the couplings from zero to $J_1^{f} = 1.0$ and $J_2^{f} = 0.9$ during a time interval $T$, namely,
\begin{equation}
\label{jonetwo}
J_{1,2}(t) \equiv 
\begin{cases}
    (t/T) J_{1,2}^{f}, \text{  } 0 \leq t \leq T, \\ 
    J_{1,2}^{f}, \text{  } t \geq {T}.
\end{cases}
\end{equation}
The system is initialized in the ground state of $H(0)$ at $t=0$, and the dynamics is calculated by exact diagonalization using the Python package QuTiP and for systems up to $N=15$. To obtain $d_{eff}$, we obtain $\rho_I(t)$ for $t$ large enough for the averages to converge and then project it in the $H_{\infty}$ basis. 
\begin{figure}
    \centering
    \includegraphics[width=.49\textwidth]{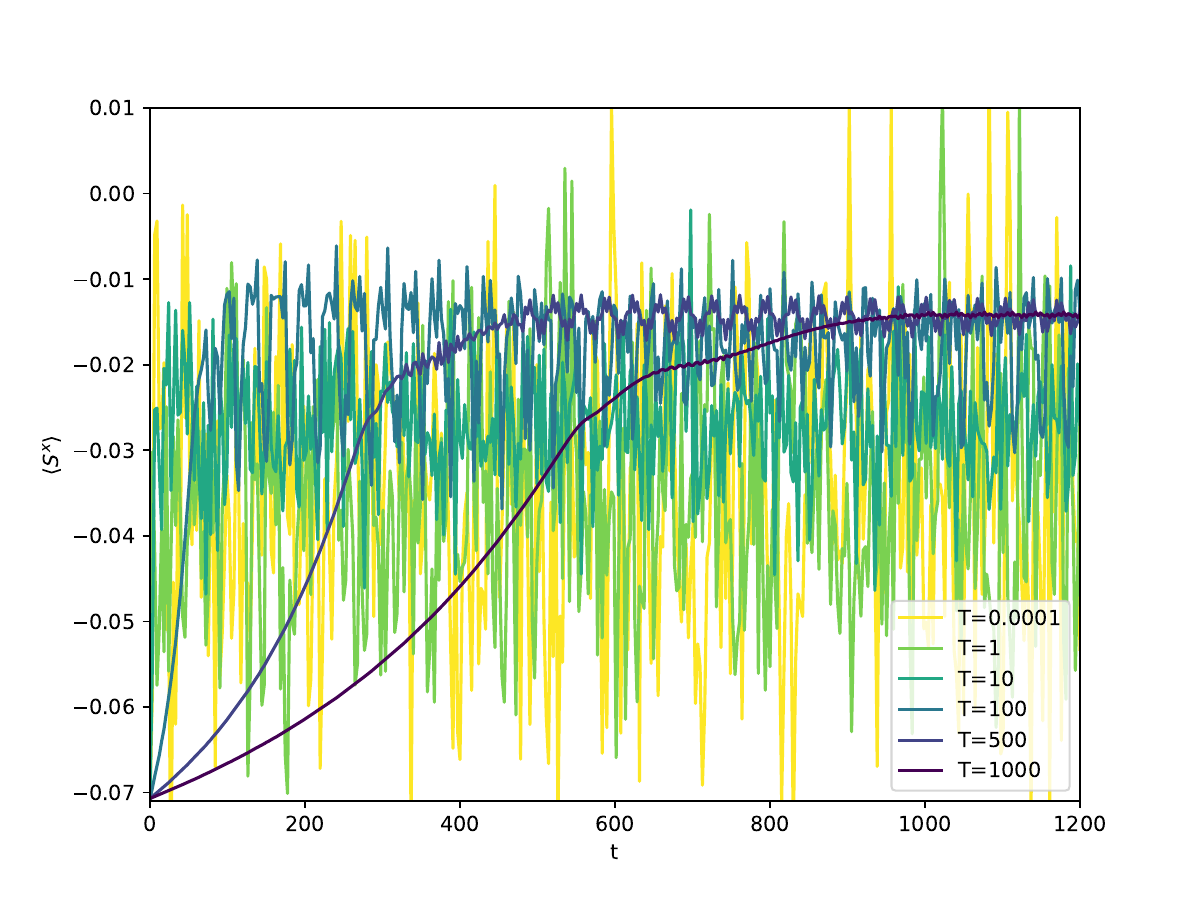}
    \caption{Time evolution of the magnetization, $\langle S^x \rangle (t)$, for an linear perturbation of duration $T$ and the parameters mentioned in the text.  It illustrates that the fluctuations decay with $T$ and that the adiabatic limit is reached for $T = 1000$. Here $N = 10$.}
    \label{fig: one}
\end{figure}

\begin{figure}
    \centering
    \includegraphics[width=.5\textwidth]{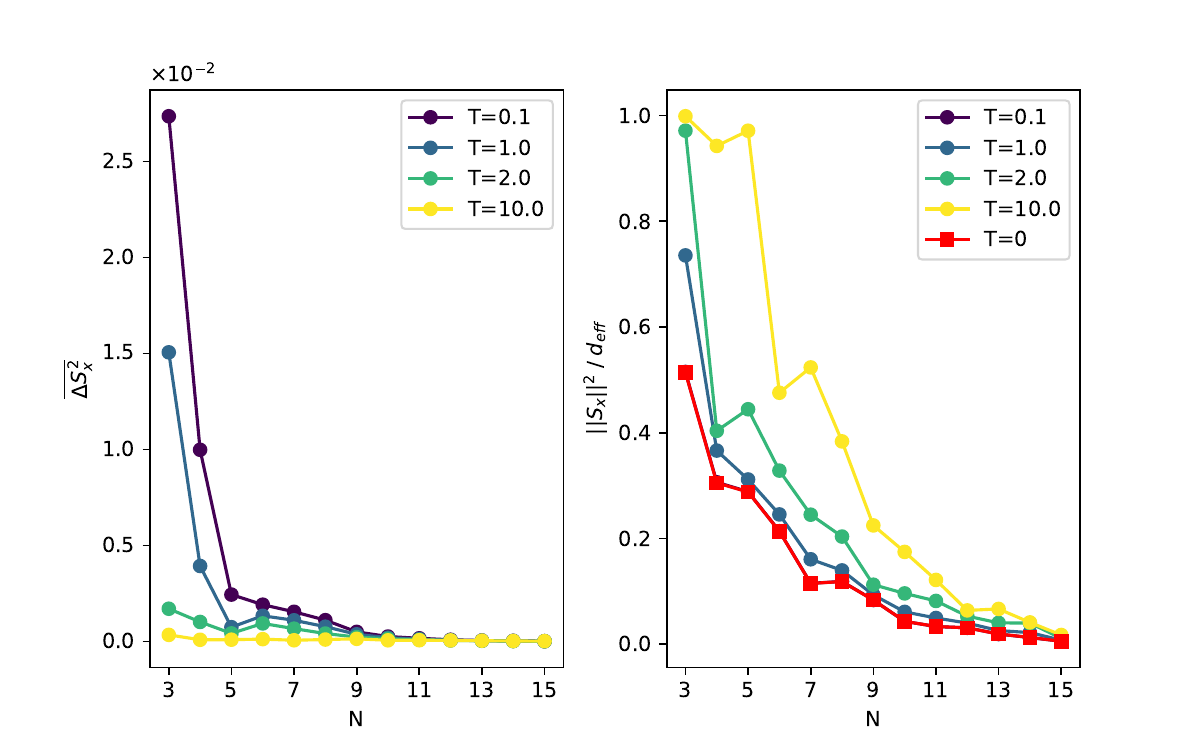}
    \caption{The average fluctuation on the magnetization, $\overline{\Delta S_{x}^2}$ (left), and the upper bound of our theorem (right) for the linear perturbation and parameter mentioned in the text. One can see that both decay with system size. We also see that for large $T$ the system is not taken out of equilibrium: the fluctuation is small even for small $N$, and the upper bound is close to 1. We checked that the curves are liner in a linear-log scale plot, indicating an exponential decay with $N$, as expected}
    \label{fig: two}
\end{figure}

Figure \ref{fig: one}, shows the magnetization in the $\hat{x}$ direction, $\langle S^x \rangle (t)$, for $N=10$ and various values of $T$, the duration of the perturbation. As $T$ increases, the fluctuations in $\langle S^x \rangle (t)$ decrease, and for sufficiently large $T$, the system enters the adiabatic regime where perturbations no longer generate significant fluctuations. 
We also checked that for $T=0.0001$, the evolution through linear interpolation is indistinguishable from that through the quench (in the graph scales).

In Figure. 2, we plot both the time-averaged fluctuation and the upper bound on the rhs of \eqref{fundamentalresult}. As expected, both quantities decrease with system size $N$, showing that equilibration will occur for macroscopic sizes. However, the values of $\overline{\Delta S_{x}^2}$ also decrease with $T$. If $T$ is large enough, then the fluctuations are negligible already for $N$ small, and therefore, there is no decay with $N$; the system never gets out of its equilibrium. This reinforces that to study equilibration, it is important to analyze the scaling of the fluctuations with $N$ and not its behavior for
a fixed system size.

On the left side, we can see the upper bound. We also see that the bound increases for finite $T$ and becomes trivial when $T$ is large. Note that the value of $T$ large enough to reach the adiabatic regime depends on $N$, which can be seen by considering the yellow curve, where the adiabatic limit is valid only for $N=3,4,5$.

\section{Conclusion}
 The physical mechanism behind macroscopic systems equilibration and thermalization has long been debated. While von Neumann, already in 1929, studied the problem in the quantum scenario and for isolated systems, many new results have appeared or been rediscovered in the last decade, particularly motivated by the ability to experimentally probe isolated quantum systems of up to dozens of particles. For the case of an instantaneous perturbation of the system, there is a rigorous theorem giving, under weak assumptions, the necessary conditions for an initial state local observables to equilibrate to the maximum entropy state: the initial state needs a large $d_{eff}$, which mean a superposition of a large number of energy eigenvalues. This is the typical situation for systems with local interactions. Thus, one can say we understand well the equilibration of isolated quantum systems since the theorems are rigorous and under weak
and reasonable assumptions.

In this work, we have extended the existing theory of quantum equilibration to account for finite-time perturbations. While the proof is a bit more intricate, the results are very similar, showing that equilibration still happens for finite-time perturbation in terms of a $d_{eff}$ that also quantifies how
strong the perturbation took the system out of equilibrium. This framework not only broadens our understanding of quantum thermalization but also opens up new avenues for exploring equilibration in driven systems or under periodic perturbations. Future work could further investigate how different perturbation protocols—such as stochastic or periodic driving—affect equilibration.

\textit{Acknowledgements. $-$} This work is supported in part the Brazilian National Institute for Quantum Information and also by funding from the Air Force Office of Scientific Research under Grant No. FA9550-23-1-0092. We acknowledge the use of ChatGPT to assist in enhancing the clarity and readability of the manuscript.

\bibliography{apssamp}

\appendix
\section{A proof of the existence of $c_{n}^{k}(\infty)$}
\label{appA}
The coefficients $c_{n}^{k}(t)$ are defined as
\begin{equation}
c_{n}^{k}(t) \equiv \langle E_{n,k} | \Psi_{I}(t) \rangle,     
\end{equation}
where $H_{\infty}|E_{n,k}\rangle = E_{n}|E_{n,k}\rangle$ and $|\Psi_{I}(t)\rangle$ is the state of the system in the interaction picture with time evolution given by $|\Psi_{I}(t)\rangle = \mathcal{U}_{I}(t,0)|\Psi_{I}(0)\rangle$. Here the interaction-picture time evolution operator $\mathcal{U}_{I}(t,\tau)$ obeys
\begin{equation}
\label{evolutionui}
i \frac{\partial \mathcal{U}_{I}(t,\tau)}{\partial t} = \delta H_{I} (t)  \mathcal{U}_{I}(t,\tau),  
\end{equation}
with $\delta H_{I} (t) \equiv e^{iH_{\infty}t} \delta H(t) e^{-iH_{\infty}t}$. In other words, $|\Psi_{I}(t)\rangle$ represents the non-trivial part of the system's dynamics, which cannot be solved exactly. 

Usual properties of the norm furnish 
\begin{multline}
\label{proof1}
|c_{n}^{k}(\tau) - c_{n}^{k}(t)| = |\langle E_{n,k}|\Big{(}\mathcal{U}_{I}(\tau,t) -\mathbf{1}\Big{)}\mathcal{U}_{I}(t,0)|E_{m}^{i}\rangle| \\ \leq ||\mathcal{U}_{I}(\tau,t) -\mathbf{1}||,   
\end{multline}
since $\| \mathcal{U}_{I}(t,0) \| = 1$.
We take the norm of the Dyson series for $\mathcal{U}_{I}(\tau,t) -\mathbf{1}$, that is,
\begin{multline}
\label{proof2}
||\mathcal{U}_{I}(\tau,t) -\mathbf{1}|| = \\ \Bigg{\|}\mathcal{T} \Bigg{(} \sum_{n=1}^{\infty} \frac{(-i)^n}{n!} \int^{\tau}_{t} dt_{1} \delta H_{I}(t_{1}) ... \int_{t}^{\tau}dt_{n} \delta H_{I}(t_{n}) \Bigg{)}\Bigg{\|},    
\end{multline}
where we have considered $\tau \geq t \geq t^{*}$.

The rhs of \eqref{proof2} can be written as
\begin{multline}
\label{proof3}
 \Bigg{\|}\mathcal{T} \Bigg{(} \sum_{n=1}^{\infty} \frac{(-i)^n}{n!} \int^{\tau}_{t} dt_{1} \delta H_{I}(t_{1}) ... \int_{t}^{\tau}dt_{n} \delta H_{I}(t_{n}) \Bigg{)}\Bigg{\|} \leq \\ \sum_{n=1}^{\infty} \frac{1}{n!} \Bigg{\|}\int^{\tau}_{t} dt_{1} \delta H_{I}(t_{1}) ... \int_{t}^{\tau}dt_{n} \delta H_{I}(t_{n})\Bigg{\|},
\end{multline}
which follows from the triangle inequality. 

The product of integrals on the rhs of \eqref{proof3} can be written as 
\begin{multline}
\label{proof4}
\Bigg{\|}\int^{\tau}_{t} dt_{1} \delta H_{I}(t_{1}) ... \int_{t}^{\tau}dt_{n} \delta H_{I}(t_{n})\Bigg{\|} \leq \\ \Bigg{\|}\int^{\tau}_{t} dt_{1} \delta H_{I}(t_{1}) \Bigg{\|} ... \Bigg{\|}\int_{t}^{\tau}dt_{n} \delta H_{I}(t_{n})\Bigg{\|},
\end{multline}
which follows from sub multiplicativity of the norm. 

Each norm which appears on rhs of \eqref{proof4} can be written as
\begin{multline}
\label{proof5}
 \Bigg{\|}\int_{t}^{\tau}dt' \delta H_{I}(t')\Bigg{\|} \leq \int_{t}^{\tau}dt' ||\delta H_{I}(t')|| \leq \\ \int_{t}^{\tau}dt' \frac{K}{(t')^{2+\epsilon}} = \frac{K}{1+\epsilon}\Bigg{(}\frac{1}{t^{1+\epsilon}}-\frac{1}{\tau^{1+\epsilon}}\Bigg{)},
\end{multline}
which follows from triangle inequality and our hypothesis that $||\delta H(t')||\leq K/(t')^{2+\epsilon}$ for $t' \geq t^{*}$.

Expressions \eqref{proof3}, \eqref{proof4} and \eqref{proof5} then lead to
\begin{multline}
\label{proof6}
\Bigg{\|}\mathcal{T} \Bigg{(} \sum_{n=1}^{\infty} \frac{(-i)^n}{n!} \int^{\tau}_{t} dt_{1} \delta H_{I}(t_{1}) ... \int_{t}^{\tau}dt_{n} \delta H_{I}(t_{n}) \Bigg{)}\Bigg{\|} \leq \\ e^{\frac{K}{1+\epsilon}\Big{(}\frac{1}{t^{1+\epsilon}}-\frac{1}{\tau^{1+\epsilon}}\Big{)}} - 1.   
\end{multline}

Finally \eqref{proof1}, \eqref{proof2} and \eqref{proof6} provide
\begin{multline}
\label{proof7}
|c_{n}^{k}(\tau) - c_{n}^{k}(t)| \leq e^{\frac{K}{1+\epsilon}\Big{(}\frac{1}{t^{1+\epsilon}}-\frac{1}{\tau^{1+\epsilon}}\Big{)}} - 1.
\end{multline}
We take the limit $\tau \rightarrow \infty$ on both sides of \eqref{proof7} to obtain
\begin{equation}
\label{proof8}
|c_{n}^{k}(\infty) - c_{n}^{k}(t)| \leq e^{\frac{K}{(1+\epsilon)t^{1+\epsilon}}} - 1,    
\end{equation}
which proves the existence of the limit $c_{n}^{k}(\infty)$. We stress that \eqref{proof8} only applies for $t \geq t^{*}$. 
\section{ Calculation of $\bar{\rho}_{I}$}
\label{appB}
We have defined the interaction-picture average state $\bar{\rho}_{I}$ as 
\begin{equation}
\label{proof9}
\bar{\rho}_{I} =  \overline{e^{iH_{\infty}t}|\Psi(t)\rangle\langle \Psi(t)|e^{-iH_{\infty}t}}.
\end{equation}
We can obtain this average value in terms of the $\{ |E_{n,k}\rangle \}$ basis, that is,
\begin{equation}
\label{proof10}
\bar{\rho}_{I} =  \sum_{m,l}\sum_{n,k} |E_{m,l}\rangle \langle E_{n,k}| \overline{c_{m}^{l}(t)(c_{n}^{k}(t))^{*}e^{i(E_{m}-E_{n})t}}.   
\end{equation}
Now we define 
\begin{equation}
\label{proof11}
\delta c_{n}^{k}(t) \equiv c_{n}^{k}(t) - c_{n}^{k}(\infty),
\end{equation}
which is useful to write 
\begin{equation}
\label{proof12}
c_{m}^{l}(t) = c_{m}^{l}(\infty) + \delta c_{m}^{l}(t)    
\end{equation}
and
\begin{equation}
\label{proof13}
c_{n}^{k}(t) = c_{n}^{k}(\infty) + \delta c_{n}^{k}(t).
\end{equation}
We substitute \eqref{proof12} and \eqref{proof13} in \eqref{proof10} to obtain 
\begin{multline}
\label{proof14}
\bar{\rho}_{I} = \\  \sum_{m,l}\sum_{n,k} |E_{m,l}\rangle \langle E_{n,k}| \overline{e^{i(E_{m}-E_{n})t}c_{m}^{l}(\infty)(c_{n}^{k}(\infty))^{*}} +\\  \sum_{m,l}\sum_{n,k} |E_{m,l}\rangle \langle E_{n,k}| \overline{e^{i(E_{m}-E_{n})t} \delta c_{m}^{l}(t)(c_{n}^{k}(\infty))^{*}} +\\  \sum_{m,l}\sum_{n,k} |E_{m,l}\rangle \langle E_{n,k}| \overline{e^{i(E_{m}-E_{n})t}c_{m}^{l}(\infty)(\delta c_{n}^{k}(t))^{*}}  +\\  \sum_{m,l}\sum_{n,k} E_{m,l}\rangle \langle E_{n,k}| \overline{e^{i(E_{m}-E_{n})t} \delta c_{m}^{l}(t)(\delta c_{n}^{k}(t))^{*}}.
\end{multline}
Each average value on the rhs of \eqref{proof14} vanishes, except for the first one. For example, the second average value leads to the calculation of
\begin{multline}
\label{proof15}
\lim_{T \rightarrow \infty} \frac{1}{T} \Bigg{(}\int_{0}^{t^{*}} dt e^{i(E_{m}-E_{n})t} \delta c_{m}^{l}(t)(c_{n}^{k}(\infty))^{*} + \\\int_{t^{*}}^{T} dt e^{i(E_{m}-E_{n})t} \delta c_{m}^{l}(t)(c_{n}^{k}(\infty))^{*} \Bigg{)},
\end{multline}
but the limit vanishes because the second integral on the rhs of \eqref{proof15} converges. Indeed,
\begin{multline}
\label{proof16}
\Bigg{|}\int_{t^{*}}^{\infty} dt e^{i(E_{m}-E_{n})t} \delta c_{m}^{l}(t)(c_{n}^{k}(\infty))^{*}\Bigg{|} \leq \\ \int_{t^{*}}^{\infty} dt |e^{i(E_{m}-E_{n})t} \delta c_{m}^{l}(t)(c_{n}^{k}(\infty))^{*}| \leq \\ \int_{t^{*}}^{\infty}dt \big{(}e^{\frac{K}{(1+\epsilon)t^{1+\epsilon}}} - 1\big{)},
\end{multline}
where we have considered that $|e^{i(E_{m}-E_{n})t}| = 1$,  $|(c_{n}^{k}(\infty))^{*}| \leq 1$ and \eqref{proof8}. The last integral in \eqref{proof16} converges because $\epsilon > 0$ by assumption. This last point follows immediately by integrating the power series for the integrand term-by-term and verifying that the resultant series converges for $\epsilon > 0$.

For the first term on the rhs of \eqref{proof14}, we choose the basis $\{ |E_{n,k}\rangle \}$ such that $|\Psi_{I}(\infty)\rangle$ has non-zero
overlap with only one eigenstate $|E_{n,k}\rangle$ (which we denote simply by $|E_{n}\rangle$) for each energy eigenvalue $E_{n}$. Thus \eqref{proof14} becomes
\begin{equation}
\label{proof17}
\bar{\rho}_{I} =  \sum_{m}\sum_{n} |E_m\rangle \langle E_n|c_{m}(\infty)(c_{n}(\infty))^{*}  \overline{e^{i(E_{m}-E_{n})t}},
\end{equation}
which provides
\begin{equation}
\label{proof18}
\bar{\rho}_{I} =  \sum_{n} |E_n\rangle \langle E_n||c_{n}(\infty)|^2,    
\end{equation}
because $\overline{e^{i(E_{m}-E_{n})t}} =  \delta_{mn}$.
\section{Proof of the Statement $\overline{\Delta A(t)^2} \leq \frac{\| A \|^2}{d_{eff}}$}
\label{AppC}
We write
\begin{equation}
 \label{proof19}
 \langle A \rangle (t) = \sum_{n,k}\sum_{m,l} (c_{n}^{k}(t))^{*}c_{m}^{l}(t) A_{nm}^{kl} e^{i(E_{n}-E_{m})t},
\end{equation}
with the definition
\begin{equation}
\label{proof20}
A_{nm}^{kl} \equiv \langle E_{n,k} | A |E_{m,l} \rangle.
\end{equation}

We rewrite \eqref{proof19} in terms of \eqref{proof11}, \eqref{proof12} and \eqref{proof13} to obtain
\begin{multline}
\label{proof21}
 \langle A \rangle (t) = \sum_{n,k}\sum_{m,l} \bigg{(}(c_{n}^{k}(\infty))^{*}c_{m}^{l}(\infty) + \zeta_{nm}^{kl}(t)\bigg{)} \times \\ A_{nm}^{kl} e^{i(E_{n}-E_{m})t},
\end{multline}
where
\begin{multline}
\label{proof22}
\zeta_{nm}^{kl}(t) \equiv  (c_{n}^{k}(\infty))^{*} \delta c_{m}^{l}(t) +  \\ (\delta c_{n}^{k}(t))^{*}c_{m}^{l}(\infty) +  (\delta c_{n}^{k}(t))^{*}\delta c_{m}^{l}(t).
\end{multline}

We can also write
\begin{equation}
\label{proof23}
\bar{A} \equiv tr (\bar{\rho}_{I} A) = \sum_{n}\sum_{k,l} (c_{n}^{k}(\infty))^{*} c_{n}^{l}(\infty) A_{nn}^{kl},
\end{equation}
where we have considered $\bar{\rho}_{I}$ as obtained in \eqref{proof18} and the choice of basis aforementioned. Thus the sum over $k$ in \eqref{proof23} has only one coefficient $c_{n}^{k}(\infty)$ different from zero for each value of $n$ and the same holds for the sum over $l$. Indeed, the expression \eqref{proof23} is completely equivalent to
\begin{equation}
\label{proof24}
\bar{A} = tr (\bar{\rho}_{I} A) = \sum_{n} |c_{n}(\infty)|^{2} A_{nn},
\end{equation}
where $A_{nn} \equiv \langle E_n| A | E_n \rangle$. 

Expressions \eqref{proof21} and \eqref{proof23} then furnish
\begin{multline}
\label{proof25}
 \langle A \rangle (t) - \bar{A} = \sum_{n \neq m}\sum_{k,l} (c_{n}^{k}(\infty))^{*}c_{m}^{l}(\infty)A_{nm}^{kl} e^{i(E_{n}-E_{m})t} \\ + \sum_{n,m}\sum_{k,l}\zeta_{nm}^{kl}(t) A_{nm}^{kl} e^{i(E_{n}-E_{m})t}
\end{multline}
and \eqref{proof25} leads to
\begin{multline}
\label{proof26}
 \overline{\Delta A (t)^2}   = \overline{\big{(}\langle A\rangle (t) - \bar{A}\big{)}^{*}\big{(}\langle A \rangle (t) - \bar{A}\big{)}} = \\ \sum_{n \neq m} \sum_{k,l}\sum_{p \neq q}\sum_{r,s} \overline{\big{(}(c_{p}^{r}(\infty))^{*}c_{q}^{s}(\infty)A_{pq}^{rs}\big{)}^{*} \times} \\ \overline{\times \big{(}(c_{n}^{k}(\infty))^{*}c_{m}^{l}(\infty)A_{nm}^{kl}\big{)} e^{i[(E_{n}-E_{m})-(E_{p}-E_{q})]t}} + \\ \sum_{n,m} \sum_{k,l}\sum_{p \neq q}\sum_{r,s}\overline{\big{(}(c_{p}^{r}(\infty))^{*}c_{q}^{s}(\infty)A_{pq}^{rs}\big{)}^{*} \times } \\ \overline{\times\big{(}\zeta_{nm}^{kl}(t)A_{nm}^{kl}\big{)} e^{i[(E_{n}-E_{m})-(E_{p}-E_{q})]t}} + \\ \sum_{n \neq m} \sum_{k,l}\sum_{p,q}\sum_{r,s}\overline{\big{(}(c_{n}^{k}(\infty))^{*}c_{m}^{l}(\infty)A_{nm}^{kl}\big{)} \times}  \\ \overline{\times\big{(}\zeta_{pq}^{rs}(t)A_{pq}^{rs}\big{)}^{*} e^{i[(E_{n}-E_{m})-(E_{p}-E_{q})]t}} + \\  \sum_{n,m} \sum_{k,l}\sum_{p,q}\sum_{r,s}\overline{\big{(}\zeta_{nm}^{kl}(t)A_{nm}^{kl}\big{)} \big{(}\zeta_{pq}^{rs}(t)A_{pq}^{rs}\big{)}^{*} \times} \\ \overline{\times e^{i[(E_{n}-E_{m})-(E_{p}-E_{q})]t}}.
\end{multline}
Each average value on the rhs of \eqref{proof26} vanishes, except for the first one. For example, the last term in \eqref{proof26} leads to the calculation of
\begin{multline}
\label{proof27}
\lim_{T \rightarrow \infty} \frac{1}{T} \Bigg{(}\int_{0}^{t^{*}} dt \big{(}\zeta_{nm}^{kl}(t)A_{nm}^{kl}\big{)}\big{(}\zeta_{pq}^{rs}(t)A_{pq}^{rs}\big{)}^{*}\times \\  \times e^{i[(E_{n}-E_{m})-(E_{p}-E_{q})]t} \\ + \int_{t^{*}}^{T} dt \big{(}\zeta_{nm}^{kl}(t)A_{nm}^{kl}\big{)}\big{(}\zeta_{pq}^{rs}(t)A_{pq}^{rs}\big{)}^{*} \times \\ \times e^{i[(E_{n}-E_{m})-(E_{p}-E_{q})]t} \Bigg{)},
\end{multline}
which vanishes, because the second integral on rhs of \eqref{proof27} converges. Indeed,
\begin{multline}
\label{proof28}
\Bigg{|}\int_{t^{*}}^{\infty} dt \big{(}\zeta_{nm}^{kl}(t)A_{nm}^{kl}\big{)}\big{(}\zeta_{pq}^{rs}(t)A_{pq}^{rs}\big{)}^{*} e^{i[(E_{n}-E_{m})-(E_{p}-E_{q})]t} \Bigg{|} \\ \leq \int_{t^{*}}^{\infty} dt |\big{(}\zeta_{nm}^{kl}(t)A_{nm}^{kl}\big{)}\big{(}\zeta_{pq}^{rs}(t)A_{pq}^{rs}\big{)}^{*} e^{i[(E_{n}-E_{m})-(E_{p}-E_{q})]t}|\\ \leq \| A \|^2 \int_{t^{*}}^{\infty} dt |\zeta_{nm}^{kl}(t)\zeta_{pq}^{rs}(t)|,
\end{multline}
where we have considered that $|e^{i[(E_{n}-E_{m})-(E_{p}-E_{q})]t}| = 1$ and $|A_{nm}^{kl}| \leq \| A \|$ (We have assumed that $A$ is bounded). 
Now the absolute value of $\zeta_{nm}^{kl}(t)$, which was defined in \eqref{proof22}, can be written as
\begin{multline}
\label{proof29}
|\zeta_{nm}^{kl}(t)| \leq |(c_{n}^{k}(\infty))^{*} \delta c_{m}^{l}(t) +  (\delta c_{n}^{k}(t))^{*}c_{m}^{l}(\infty)  + \\  (\delta c_{n}^{k}(t))^{*}\delta c_{m}^{l}(t)| \leq |(c_{n}^{k}(\infty))^{*} \delta c_{m}^{l}(t)| +  |(\delta c_{n}^{k}(t))^{*}c_{m}^{l}(\infty)| \\ + |(\delta c_{n}^{k}(t))^{*}\delta c_{m}^{l}(t)|, 
\end{multline}
which follows from triangle inequality. If we consider that $|c_{n}^{k}(\infty)| \leq 1$ and \eqref{proof8}, then \eqref{proof29} leads to
\begin{equation}
\label{proof30}
|\zeta_{nm}^{kl}(t)| \leq e^{\frac{2K}{(1+\epsilon) t^{1+\epsilon}}} - 1.
\end{equation}
Inequalities \eqref{proof28} and \eqref{proof30} then lead to
\begin{multline}
\label{proof31}
\Bigg{|}\int_{t^{*}}^{\infty} dt \big{(}\zeta_{nm}^{kl}(t)A_{nm}^{kl}\big{)}\big{(}\zeta_{pq}^{rs}(t)A_{pq}^{rs}\big{)}^{*} e^{i[(E_{n}-E_{m})-(E_{p}-E_{q})]t} \Bigg{|}  \\ \leq \| A \|^2 \int_{t^{*}}^{\infty} dt \big{(}e^{\frac{2K}{(1+\epsilon) t^{1+\epsilon}}} - 1\big{)}^2
\end{multline}
and the last integral converges for $\epsilon > 0$.

The first average value in \eqref{proof26} then provides
\begin{multline}
\label{proof32}
 \overline{\Delta A (t)^{2}}  = \sum_{n \neq m} \sum_{p \neq q}\big{(}(c_{p}(\infty))^{*}c_{q}(\infty)A_{pq}\big{)}^{*} \times \\ \times\big{(}(c_{n}(\infty))^{*}c_{m}(\infty)A_{nm}\big{)} \overline{e^{i[(E_{n}-E_{m})-(E_{p}-E_{q})]t}},
\end{multline}
which follows from the choice of basis aforementioned. 

We denote
\begin{equation}
\label{proof33} 
v_{\alpha} \equiv v_{(p,q)} \equiv (c_{p}(\infty))^{*}c_{q}(\infty)A_{pq},
\end{equation}
\begin{equation}
\label{proof34}
v_{\beta} \equiv v_{(n,m)} \equiv (c_{n}(\infty))^{*}c_{m}(\infty)A_{nm},
\end{equation}
\begin{equation}
\label{proof35}
G_{\alpha} \equiv E_{p} - E_{q}
\end{equation}
and
\begin{equation}
\label{proof36}
G_{\beta} \equiv E_{n} - E_{m}.
\end{equation}
The expression \eqref{proof32} can then be written as 
\begin{multline}
\label{proof37}
\overline{\Delta A(t)^2} = \lim_{T \rightarrow \infty} \frac{1}{T} \int_{0}^{T} dt \sum_{\alpha,\beta} v_{\alpha}^{*}v_{\beta} e^{i(G_{\beta}-G_{\alpha})t} = \\ = \sum_{\alpha}|v_{\alpha}|^2,
\end{multline}
since 
\begin{equation}
\label{proof38}
\lim_{T \rightarrow \infty} \frac{1}{T} \int_{0}^{T} dt  e^{i(G_{\beta}-G_{\alpha})t} = \delta_{\alpha \beta}.
\end{equation}
The preceding equality follows from the hypothesis that there are no energy gap degeneracies.

We rewrite \eqref{proof37} as
\begin{multline}
\label{proof39}
\overline{\Delta A(t)^2} = \sum_{p \neq q} |(c_{p}(\infty))^{*}c_{q}(\infty)A_{pq}|^2 \leq \\ \sum_{p,q} |(c_{p}(\infty))^{*}c_{q}(\infty)A_{pq}|^2 = tr (A \bar{\rho}_{I}A^{\dagger}\bar{\rho}_{I}) \leq tr(A^2 \bar{\rho}_{I}^2).
\end{multline}
The last inequality follows from Schwartz inequality, namely,
\begin{equation}
\label{proof40}
|tr(\xi^{\dagger}\eta)| \leq \sqrt{tr(\xi^{\dagger} \xi)} \sqrt{tr(\eta^{\dagger}\eta)} 
\end{equation}
with $\xi \rightarrow \bar{\rho}_{I}A^{\dagger}$ and $\eta \rightarrow A^{\dagger} \bar{\rho}_{I}$. 

Finally, for two positive operators $C$ and $D$ 
\begin{equation}
\label{proof41}
tr(CD) \leq \| C \| tr(D).
\end{equation}
If we consider $C \rightarrow A^2 = A^{\dagger}A$ and $D \rightarrow \bar{\rho}_{I}^{2}$, then \eqref{proof39} and \eqref{proof41} lead to
\begin{equation}
\label{proof42}
\overline{\Delta A(t)^2} \leq \frac{\| A \|^2}{d_{eff}},
\end{equation}
where $d_{eff} \equiv 1/tr(\bar{\rho}_{I}^2)$.
\section{A proof that $\overline{A-\bar{A}} = 0$}
\label{AppD}
We have defined $\bar{A} \equiv tr(\rho_{I}A)$ and now we prove that this quantity actually is the same as $\overline{\langle A \rangle (t)}$. We can take the infinite-time average on both sides of \eqref{proof25}, that is,
\begin{multline}
\label{proof43}
 \overline{\langle A \rangle (t) - \bar{A}} = \sum_{n \neq m}\sum_{k,l} (c_{n}^{k}(\infty))^{*}c_{m}^{l}(\infty)A_{nm}^{kl} \overline{e^{i(E_{n}-E_{m})t}} + \\ + \sum_{n,m}\sum_{k,l}\overline{\zeta_{nm}^{kl}(t) A_{nm}^{kl} e^{i(E_{n}-E_{m})t}}.
\end{multline}
The first average value on the rhs of \eqref{proof43} provides $\overline{e^{i(E_{n}-E_{m})t}} = \delta_{nm}$ and the first sum vanishes, because $n \neq m$. The second sum on the rhs of \eqref{proof43} also vanishes, because the average inside it vanishes. In order to clarify this last point, we write
\begin{multline}
\label{proof44}
\overline{\zeta_{nm}^{kl}(t) A_{nm}^{kl} e^{i(E_{n}-E_{m})t}} = \\ \lim_{T \rightarrow \infty} \frac{1}{T} \Bigg{(} \int_{0}^{t^*}dt \zeta_{nm}^{kl}(t) A_{nm}^{kl} e^{i(E_{n}-E_{m})t} + \\ + \int_{t^{*}}^{T}dt \zeta_{nm}^{kl}(t) A_{nm}^{kl} e^{i(E_{n}-E_{m})t} \Bigg{)},
\end{multline}
which vanishes, because the second integral on the rhs of \eqref{proof44} converges. Indeed,
\begin{multline}
\label{proof45}
\Bigg{|}  \int_{t^{*}}^{\infty}dt \zeta_{nm}^{kl}(t) A_{nm}^{kl} e^{i(E_{n}-E_{m})t}  \Bigg{|} \leq  \int_{t^{*}}^{\infty}dt|\zeta_{nm}^{kl}(t)| |A_{nm}^{kl}| \\ \leq \| A \| \int_{t^{*}}^{\infty}dt\big{(}e^{\frac{2K}{(1+\epsilon) t^{1+\epsilon}}} - 1\big{)},
\end{multline}
which converges for $\epsilon >  0$. The last inequality on the rhs of \eqref{proof45} follows from \eqref{proof30} and from the fact that $A$ is bounded.
\section{Quench of the Hamiltonian as a Particular Case of the Theorem \eqref{proof42}}
\label{AppE}
We consider the Hamiltonian
$H(t) = H_{0} + \lambda(t) \Delta H$ with $\lambda(t)$ defined as 
\begin{equation}
\label{proof46}
\lambda(t) \equiv 
\begin{cases}
    t/t^{*}, \text{  } 0 \leq t \leq t^{*}, \\ 
    1, \text{  } t \geq t^{*}.
\end{cases}
\end{equation}
Here $\delta H(t) = (\lambda(t) - 1)$ and $\delta H(t) \equiv 0$ (for $t \geq t^{*}$) by assumption.

For the quench case, we write
 \begin{multline}
 \label{proof47}
\bar{\rho}_{I} \equiv \overline{e^{iH_{\infty} t} \rho(t) e^{-iH_{\infty}t}} = \\ = \lim_{t^{*} \rightarrow 0^{+}}\Bigg{(}\lim_{T \rightarrow \infty} \frac{1}{T} \int_{0}^{T} e^{iH_{\infty} t} \rho(t) e^{-iH_{\infty}t} dt \Bigg{)}.
 \end{multline}
(The symbol plus denotes hand-right limit) We then consider the result in \eqref{proof18}, namely,
\begin{equation}
\label{proof48} 
\lim_{T \rightarrow \infty} \frac{1}{T} \int_{0}^{T} e^{iH_{\infty} t} \rho(t) e^{-iH_{\infty}t} dt = \sum_{n} |c_{n}(\infty)|^2|E_{n}\rangle \langle E_{n}|.
\end{equation}
But
\begin{multline}
\label{proof49}
c_{n}(\infty) = \langle E_{n}|\mathcal{U}_{I}(\infty,t^{*})\mathcal{U}_{I}(t^{*},0)|E_{m}^i\rangle = \\ = \langle E_{n}|\mathcal{U}_{I}(t^{*},0)|E_{m}^i\rangle = c_{n}(t^{*}),
\end{multline}
because $\mathcal{U}_{I}(\infty,t^{*}) = \mathbf{1}$. This last equality holds, because $\delta H_{I}(t) \equiv e^{iH_{\infty}t} \delta H(t)e^{-iH_{\infty}t} \equiv 0$ (for $t \geq t^{*}$) by assumption. It follows from \eqref{proof48} and \eqref{proof49} that    
\begin{equation}
\label{proof50} 
\lim_{T \rightarrow \infty} \frac{1}{T} \int_{0}^{T} e^{iH_{\infty} t} \rho(t) e^{-iH_{\infty}t} dt = \sum_{n} |c_{n}(t^{*})|^2 |E_{n}\rangle \langle E_{n}|.
\end{equation}
We remember that $c_{n}(t) \equiv e^{iE_{n}t}d_{n}(t)$, then \eqref{proof47} and \eqref{proof50} furnish
\begin{multline}
\label{proof51}
 \bar{\rho}_{I} \equiv \overline{e^{iH_{\infty} t} \rho(t) e^{-iH_{\infty}t}} = \lim_{t^{*} \rightarrow 0+} \sum_{n} |d_{n}(t^{*})|^2 |E_{n}\rangle \langle E_{n}|\\ = \sum_{n} |d_{n}(0^{+})|^2 |E_{n}\rangle \langle E_{n}|.
\end{multline}
    It follows from \eqref{proof51} that $tr(\bar{\rho}_{I}^2) = tr(\bar{\rho}^{2})$ and our result in \eqref{fundamentalresult} reduces to that in \eqref{fundamentalquench}.

On the other hand, if we consider the limits in \eqref{proof47} in reverse order, then we obtain 
 \begin{multline}
 \label{proof52}
\bar{\rho}_{I} \equiv \overline{e^{iH_{\infty} t} \rho(t) e^{-iH_{\infty}t}} = \\ = \lim_{T \rightarrow \infty} \Bigg{(}\lim_{t^{*} \rightarrow 0} \frac{1}{T} \int_{0}^{T} e^{iH_{\infty} t} |\Psi(t)\rangle \langle \Psi(t)| e^{-iH_{\infty}t} dt \Bigg{)} \\
= \lim_{T \rightarrow \infty} \frac{1}{T} \int_{0}^{T} |\Psi(0)\rangle \langle \Psi(0)| dt = \rho(0)
 \end{multline}
and $tr(\bar{\rho}_{I}^2) = tr(\rho(0)^{2}) = 1$. However, we observe that this last procedure is not correct, because the quench case
is appropriately treated by considering a finite change of the Hamiltonian, i.e., one that occurs during a finite time $t^{*}$, and then taking the limit $t^{*} \rightarrow 0^{+}$. The results obtained in \eqref{proof51} and \eqref{proof52} are different, because the limits do not commute. However, a numerical simulation (like ours above) can confirm that \eqref{proof47} leads to the correct effective dimension $d_{eff}$ for stepwise $H(t)$. 
\end{document}